\documentclass[review, onecolumn,english,american,prl,manuscript,superscriptaddress]{revtex4}
\usepackage[LGR,T1]{fontenc}
\usepackage[latin9]{inputenc}
\usepackage{color}
\usepackage{multirow}
\usepackage{amsmath}
\usepackage{amssymb}
\usepackage{graphicx}
\usepackage{textcomp}

\usepackage{array}
\newcolumntype{L}{>{\centering\arraybackslash}m{2cm}}

\makeatletter

\ProvideTextCommand{\~}{LGR}[1]{\char126#1}

\@ifundefined{textcolor}{}
{%
 \definecolor{BLACK}{gray}{0}
 \definecolor{WHITE}{gray}{1}
 \definecolor{RED}{rgb}{1,0,0}
 \definecolor{GREEN}{rgb}{0,1,0}
 \definecolor{BLUE}{rgb}{0,0,1}
 \definecolor{CYAN}{cmyk}{1,0,0,0}
 \definecolor{MAGENTA}{cmyk}{0,1,0,0}
 \definecolor{YELLOW}{cmyk}{0,0,1,0}
}

\usepackage[english]{babel}

\def\url#1{}
\addto\captionsenglish{%
}

\usepackage[none]{hyphenat}
\usepackage{multirow}

\usepackage{babel}

\makeatother

\usepackage{babel}

\usepackage{hyperref}
\usepackage{xcolor}

\begin{document}

\title{Machine-Learning-Enabled Fast Optical Identification and Characterization of 2D Materials}
\author{Polina A. Leger}
\author{Aditya Ramesh}
\author{Talianna Ulloa}
\affiliation{Department of Electrical and Computer Engineering, University of Florida, Gainesville, Florida 32611, USA.}
\author{Yingying Wu}
\affiliation{Department of Electrical and Computer Engineering, University of
Florida, Gainesville, Florida 32611,
USA.}
\thanks{Correspond to: yingyingwu@ufl.edu}
%author list can be discussed later in our group meeting

\begin{abstract}
Two-dimensional materials are a class of atomically thin materials with assorted electronic and quantum properties. Accurate identification of layer thickness, especially for a single monolayer, is crucial for their characterization. This characterization process, however, is often time-consuming, requiring highly skilled researchers and expensive equipment like atomic force microscopy. This project aims to streamline the identification process by using machine learning to analyze optical images and quickly determine layer thickness. In this paper, we evaluate the performance of three machine learning models - SegNet, 1D U-Net, and 2D U-Net- in accurately identifying monolayers in microscopic images. Additionally, we explore labeling and image processing techniques to determine the most effective method for identifying layer thickness in this class of materials. 
\end{abstract}
\maketitle

\newpage

\section{Introduction} 
Two-dimensional (2D) van der Waals (vdW) materials are a class of materials that can be atomically thin, down to a thickness of $\sim$0.8 nm. These materials exhibit weak interlayer vdW interactions and strong intralayer covalent bonding. Due to their unique interlayer interaction, they provide access to a broader class of materials with tunable properties such as energy dispersion relations, bandgap control and carrier mobility~\cite{han2018investigation,chen2016probing,wu2016negative,wang2020topological,zhang20242d,zhong2024integrating,wu2020large,wu2020neel}. This tunability allows for greater design control over device properties, making them suitable for applications ranging from insulators in gate dielectrics to groundbreaking new electronics, such as superconductors for quantum computers~\cite{ajayan_two-dimensional_2016,wang2020topological,wu2019induced,hou2023ubiquitous,han2018investigation}. These materials were first discovered in 2004 when graphene was fabricated using a piece of scotch 
tape to exfoliate an atoms-thick layer from graphite~\cite{gerstner_nobel_2010,geim2007rise}. Today, large-scale 
manufacturing processes, such as vapor-phase epitaxy\cite{zhang2022epitaxy,tong2019vapor,walsh2017van} and chemical vapor deposition\cite{cai2018chemical,yu2015synthesis,bhowmik2022chemical}, are employed 
alongside the exfoliation method to grow these materials. Precisely controlling variables such as layer thickness during fabrication is challenging. As a result, characterization techniques are essential to properly identifying the grown material based on its physical and electrical properties. 

Optical spectroscopy\cite{alexeev2017imaging, kenaz2023thickness}, Raman spectroscopy\cite{wu2016negative,wu2020large}, photoluminescence\cite{wu2019induced,jie2018luminescence}, and atomic force microscopy\cite{chen2015high,wu2020neel} are some of the techniques used to determine the thickness of 2D flakes. However, these methods can be time-consuming and inefficient. In addition, the delicate nature of 2D materials\cite{duong2017van,kim2021surface,wu2022manipulating} requires complex set-up like a closed environment with inert gas, making certain characterization techniques expensive and not easy to operate. Among these methods, optical microscopy stands out as an easy and cost-effective method. It works by measuring the light path's reflection at different intensities, depending on material absorption, with the light travelling an additional distance of material thickness when incident on the underlying substrate. The resulting interference between all the wavelengths of incident light can be calculated using contrast equation in Equation~\ref{contrast}, where R$_\textrm{mat}$ and R$_\textrm{sub}$ denote the reflection spectra of the 2D flake and the substrate~\cite{mao_thickness_2023}, respectively.
\begin{equation}
    C(\lambda) = \frac{R_\textrm{sub}(\lambda) - R_\textrm{mat}(\lambda)}{R_\textrm{sub}(\lambda)}
    \label{contrast}
\end{equation}
The RGB values of each pixel are the averaged contrast spectra at the corresponding range of wavelengths. This allows a functional mapping that associates the thickness of a 2D sheet with its apparent color, like in the case of graphene~\cite{mao_thickness_2023}. However, the contrast spectra vary depending on illumination, substrate thicknesses, substrate type, and materials being studied, leading to a tedious process of determining the correct functional mapping for each sample. To address these issues, machine learning can be employed to find the accurate relationship between pixel color and layer thickness, streamlining the characterization of 2D materials.

Machine learning, a subset of artificial intelligence (AI), predicts the underlying patterns in a learning data set and extracts features without manually defined functions. There are two main types of machine learning algorithms, supervised and unsupervised. Supervised learning algorithms learn to predict features in a sample using knowledge from labeled examples. They compare predictions to expected classification labels to continually refine the model~\cite{mao_thickness_2023}. Unsupervised learning algorithms, on the other hand, do not require labeled examples and refine the model by identifying groups with similar properties, such as pixel color. 
For the complex task of thickness identification of 2D materials, supervised machine learning is often preferred due to its higher accuracy and less manipulation of results~\cite{wei2019machine,morgan2020opportunities,duanyang2023application,schmidt2019recent}. However, it requires a dataset containing both the original optical images and the correct pre-classification of each pixel according to layer thickness.\,This is a significant challenge since optical images of 2D flakes do not automatically indicate layer locations- the core problem this paper aims to solve. Once a sufficient set of optical images is labeled, a trained machine learning model can automatically classify the layers.
As reviewed by Mao et al.~\cite{mao_thickness_2023}, previous studies have successfully implemented machine learning algorithms for automatic layer characterization\cite{dihingia2024quantifying} and other material properties\cite{zhu2023machine,solis2022machine,li2023thermal} such as mechanical strength\cite{sattari2020prediction}. In our work, we systematically tested various approaches to find the best generalized solution for 2D material thickness characterization. Performance of three machine learning models-SegNet, 1D U-Net and 2D U-Net is compared in accurately identifying monolayers in microscopic images.

\section{Methods} 
Given that machine learning is a fast and effective method for layer identification, we undertook a comprehensive approach to leverage its capabilities. We selected three machine learning models, acquired a dataset to train and test the models, and obtained labeled counterparts for the dataset. We then preprocessed the images, observed the results and metrics from the model prediction, and fine-tuned the models to identify the best-performing one.
%can maybe add one more figure to summarize this procedure and give some visualization to the most important point in the introduction
We tested multiple machine learning models, with a focus on object segmentation models. Various image preprocessing methods and performance metrics were utilized, and visual results were used to validate and refine the methods that best supplemented the machine learning models. The algorithms were trained on laptop computers with eight 1.80 GHz processors or the Linux workstations in the ECE computer lab at University of Florida (UF).

%\begin{figure*}[ht]
%\begin{centering}
 %\includegraphics[width=0.85\textwidth]{fig1.jpg} 
%\par\end{centering}
%\caption{Illus. }\label{Fig1} 
%\end{figure*}

\section{Preprocessing}\label{sec:preprocess}
In segmentation problems, image preprocessing is an integral step to extracting only the important features to be identified such as contrasts, shapes, or dimensionality and filter out imperfections that could confuse the machine learning algorithm. Particularly, image manipulation is used as a standalone method for layer identification but this process still requires human supervision and must be adjusted based on type of material and how the samples were acquired~\cite{gao_total_2008,  mao_thickness_2023,li2013rapid}. This section details the steps we used to identify an image preprocessing technique that would improve accuracy of the ML model.

\begin{figure}[ht]
    \centering
    %need to get graphics with grayscale image
    \includegraphics[width=16cm]{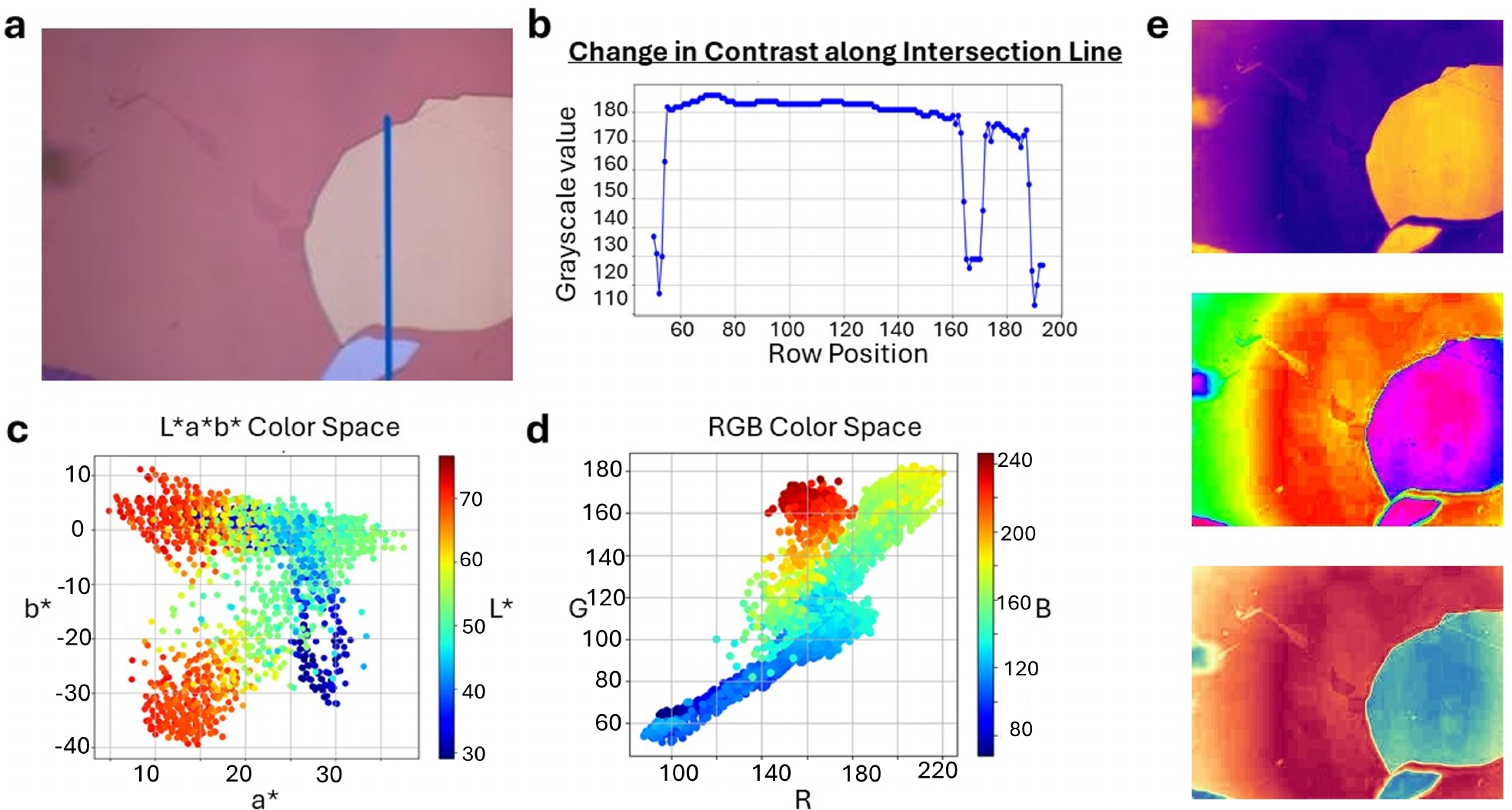}
    \vspace{-10pt}
    \caption{Image Preprocessing Techniques. (a) The original image before any preprocessing. The blue line indicates where the contrast measurements were taken from. (b) The contrast measurements along the blue line in the original image. The edges of the material are clearly demarcated by large jumps in contrast. (c) L*a*b* diagram with lightness information indicated by color and color information indicated by position: +a* for the red direction, -a* for the green direction, +b* for the yellow direction, and -b* for the blue direction. (d) RGB diagram with R and G for red and green color intensity and B for blue color represented by color of the data points. (e) Image preprocessing method total color difference (TCD) converted to various color spaces, from top to bottom: plasma, HSV and spectral color maps. }
    \label{fig:colorprocess}
\end{figure}

One of the simplest methods for image normalization is grayscale conversion which only relays lightness information of each pixel.  Due to its simplicity, we analyzed the grayscale values across an image to pick out contrast variations across substrate and material. This was accomplished by placing a line across the image and measuring the pixel-by-pixel grayscale value. These values were recorded and plotted, providing a visualization that depicted the correlation between contrast variation and changes in layer composition. As illustrated in Figs.~\ref{fig:colorprocess}a-b, discernible shifts in lightness were observed as the line crossed from the substrate into the material. However, as the line crossed over the monolayer in the image, the results were not as consistent.

The grayscale is not enough information to identify the layers of the optical images because contrast is impacted by color as well as lightness because each wavelength of the light illuminating the image contributes to the contrast~\cite{gao_total_2008}. The preferred color space in layer identification is L*a*b* space~\cite{gao_total_2008, mao_thickness_2023}. The L*a*b color space consists of three coordinates: L for lightness, representing color intensity from black to white; A, representing the difference from green to red; and B, representing the difference from blue to yellow~\cite{cielab_io}. Compared to RGB, the L*a*b* color space is more representative of what humans see and offers perceptually uniform color distinctions independent of the device as demonstrated in Figs.~\ref{fig:colorprocess}c-d~\cite{li2019rapid}. The L*a*b* representation of pixel color shows a wider spread, and the color of the data points showing Lightness overlapped (as compared to blue being restricted to specific red and green regions) meaning more unique information can be extracted from the L*a*b* space.

Using L*a*b*, we explored the technique of total color difference (TCD). TCD quantifies the difference between two colors, measuring the color difference between material and substrate. In the TCD approach, the substrate is factored in by measuring the mean L*a*b values for the substrate and subtracting from each pixel's L*a*b* value to derive a $\Delta E$ value as shown in Equation~\ref{eqn:tcd}~\cite{gao_total_2008}. 
Once normalized based on substrate values, various color maps are applied according to these derived values to allow visualization. Three different color maps, plasma, HSV, and diverging or spectral color maps, were compared as shown in Fig.~\ref{fig:colorprocess}e. These color maps were employed in testing image segmentation models to evaluate their effectiveness.
\begin{equation}
    \Delta E = (\Delta L^*)^2 + (\Delta a^*)^2 + (\Delta b^*)^2
    \label{eqn:tcd}
\end{equation}
Lastly, the~\ref{sec:label} will detail a third preprocessing method that was adapted from ~\cite{nanomaterial_deeplearning}.

\section{Labeling and Data Augmentation}\label{sec:label}
\begin{figure}[htb]
    \centering
    \includegraphics[width=16cm]{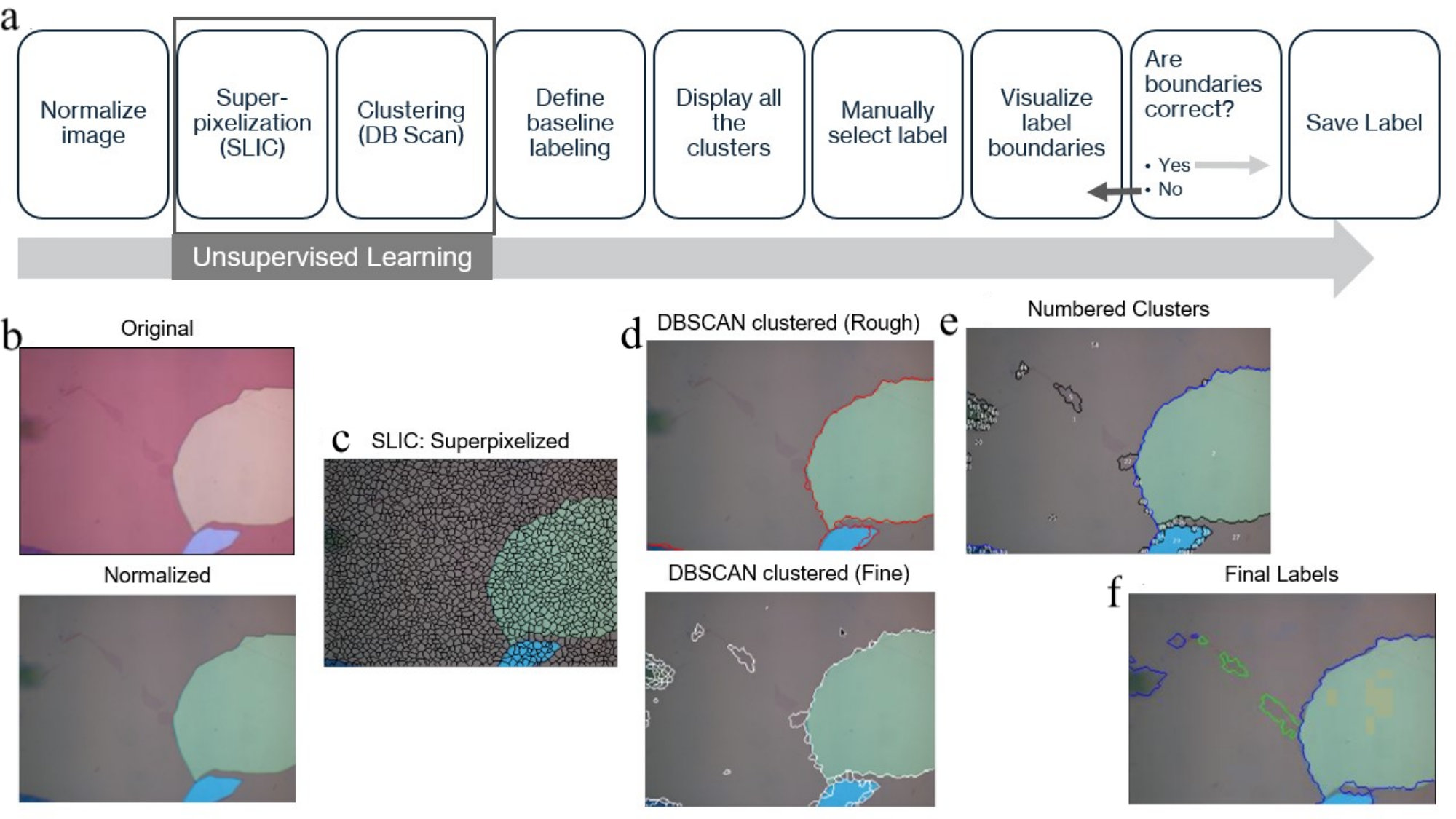}
    \vspace{-25pt}
    \caption{Pre-labeling a dataset. (a) Flowchart with detailed steps shown in (b) original and normalized image, (c) superpixelization, (d) all the clusters displayed, (e) numbered clusters and (f) final labels.}
    \label{fig:labelingprocess}
\end{figure}

As mentioned in the introduction, a dataset must initially be labeled through the tedious process of defining the boundaries of each class on a select group of training images. The labeling process is summarized in Fig.~\ref{fig:labelingprocess} and facilitated by two unsupervised learning algorithms: SLIC and DBSCAN. After acquiring a labeled dataset, the number of labeled images can be multiplied by data augmentation. This reduces overfitting resulting from insufficient diversity in a dataset and improves the generalization of results to better align with real-world scenarios~\cite{shorten2019survey}. Most of the code for labeling and data augmentation was adapted from the $Nanomaterial\_DeepLearning$ Github repository by Yafang Yang~\cite{nanomaterial_deeplearning}, which was published in~\cite{han_deeplearningenabled_2020}. These methods will be described below.

\begin{figure}[htb]
    \centering
    \includegraphics[width=16cm]{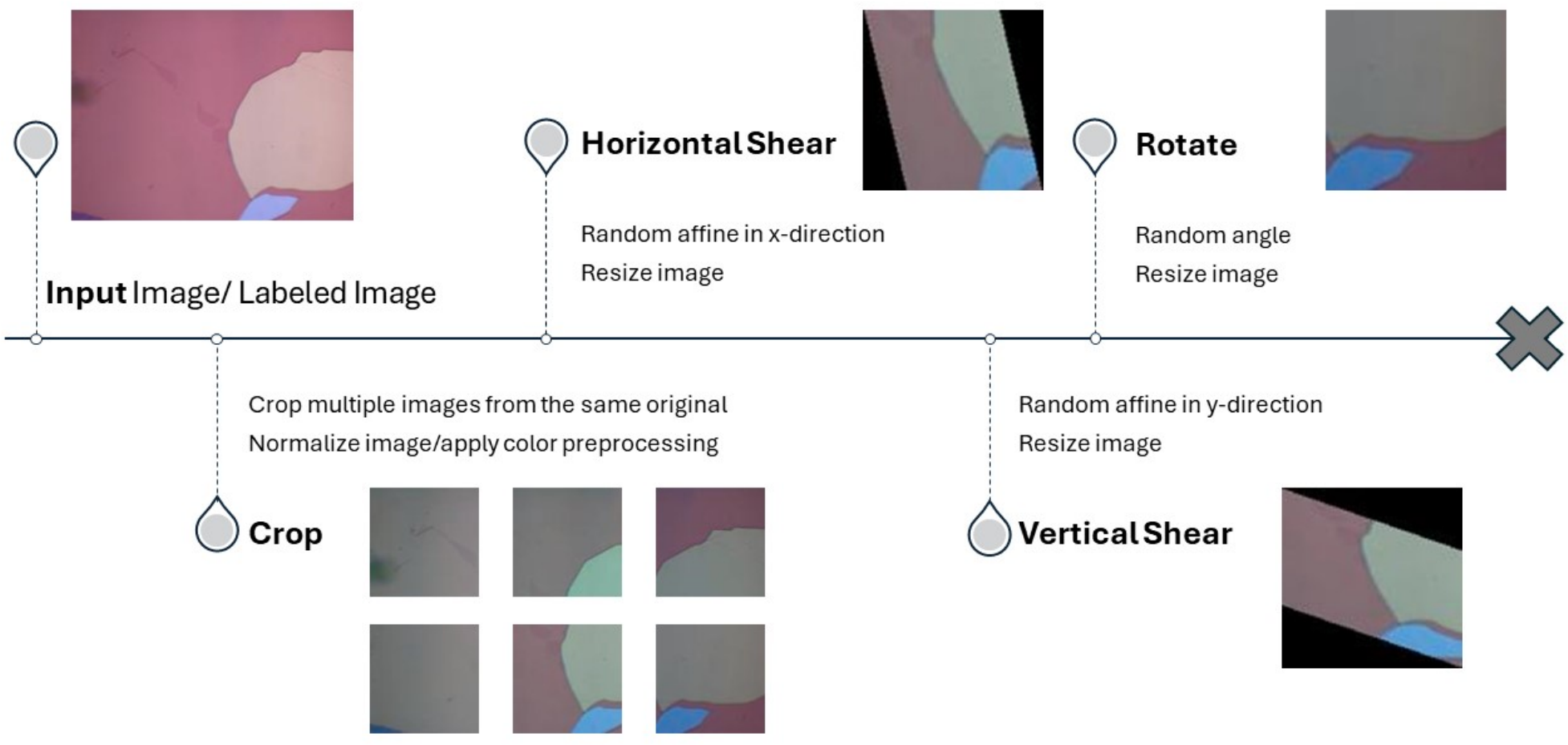}
    \vspace{-25pt}
    % NEED TO remove g from the image
    \caption{ Step-by-step data augmentation algorithm: starting from input, perform cropping, horizontal shear, vertical shear, and rotation. Note that black regions pose issues, which can potentially be addressed by introducing a new class. Hopefully, future improvements will provide a better solution.}
    \label{fig:dataaugprocess}
\end{figure}

As shown in Fig.~\ref{fig:labelingprocess}, the initial step involves normalizing the images to mitigate variations between different experimental set-ups. This is achieved by converting the image into the  L*a*b* color space as explained in Section \ref{sec:preprocess}, then adjusting it based on the image's median with the following rules: \\
Rule 1. Divide lightness of specific pixel by $\frac{L_\textrm{median}}{100} * 2 $\\
Rule 2. Subtract $a_\textrm{median}$ and $b_\textrm{median}$ from the specific pixel's a and b values. \\
Rule 3. Size image to one-tenth of it's original size. \\
The last step is so the pictures are processed faster.

Next, two unsupervised learning methods are adapted from Peter Koveski's Image Segmentation Package. The first is called SLIC and adapts K-means clustering which can create many large "superpixels" (groups of pixels that are similar) according to a distance factor, how small these superpixels can be, and how many total superpixels should be defined. This is shown in Fig.~\ref{fig:labelingprocess}c. Then the DBSCAN learning method implements spectral clustering to group the superpixels according to a distance function that factors in both the color difference and the spatial difference between superpixels to determine whether to group them into a cluster. The determining factor is a threshold that when set at two different values differentiates between the fineness of the monolayer and the roughness of the multilayer. The fine and rough DBSCAN results are shown in Fig.~\ref{fig:labelingprocess}d. This is when the multilayer is differentiated and the fine clusters, superpixel values, and an initial labeling are saved into matrices of the same size as the image. The labeling used was \{1 = substrate, 2 = monolayer, 3 = multilayer\}. The logic needed to be severely reworked in this section because the indexing was illegal, and the clusters were labeled in a confusing manner. Then the software performs a few checks to ensure all the pre-labeling is accurate before the full-sized image is displayed with the cluster boundaries and their labels as shown in Fig.~\ref{fig:labelingprocess}e.
Finally, the user is directed to label which sections should be part of the multilayer (Fig. \ref{fig:labelingprocess}f) that aren't enclosed in the blue boundaries. This was useful for samples that consisted of more multilayer than substrate and confused the machine learning algorithms. Then the user was directed to state which clusters are a part of the monolayer. This takes some expertise from the user to identify which flakes are monolayers and which aren't, but as stated before, the lightest contrast regions correspond to the thinnest layers. Finally, clusters that should be part of the background are corrected. The program displays the new hypothetical class boundaries and asks the user if they are happy with the results or want to restart. When the user stops retrying, the program saves the labels into a .csv, the original image, and the boundaries image just for the user to refer back to. If the original clusters and superpixels do not properly encompass all the necessary details in the sample, the hyperparameters for the SLIC and DBSCAN methods can be adjusted and rerun.

When the images are labeled, the labels and original images are fed into the data augmentation process shown in Fig.~\ref{fig:dataaugprocess}. This crops the images into a 100$\times$100 image. The cropping was done systematically by fitting in the rounded-up quotient of the original image size to the new cropped size by overlapping the last image with the second-to-last image. The cropping was also randomized to compare the results. Then the images were color-processed, then sheared in two directions and rotated. This resulted in four images for each crop and there was approximately six crops per image in the dataset we used of eighteen images leading to 408 total training images. The original dataset somehow had 1400 images but it seems the images are different than the ones that were designated to the labeling program. In addition, the shearing was accomplished by affine2d and an imresize in MATLAB. However, due to pixel averaging this resulted in a region of zeros where the image is sheared. This was accounted for in the machine learning program by adding a new class with the label 4. Although neither 4 or 0 could arise from the labeling program, the machine learning struggled with labeling when the first class (0) was a placeholder and not present in all images. This was compared to leaving the new classification class out and keeping only three classes.

\section{Models and Results}

Object segmentation models were surveyed and utilized to predict regions of monolayer, multilayer, and substrate. Object segmentation models utilize supervised learning to classify each pixel of an image into certain classes. In this work, two models were investigated: SegNet~\cite{badrinarayanan_segnet_2017} and U-Net~\cite{ronneberger_unet_2015}. In addition, both 1D and 2D versions of U-Net were trained to compare performance on physics-based features versus image-based features respectively. The 1D U-Net model was motivated based on linear dependence of light transmittance on the number of layers\cite{niu2018thickness}. In addition, since each row is extracted for each image, there are more samples for training a smaller model when compared to the 2D version, allowing for potentially better generalization and predictions. 

Since this is a multi-class classification problem, cross-entropy loss was utilized as shown in Equation~\ref{eqn:crossentropy}. Cross-entropy measures the degree of randomness between the true labels ($T_{ni}$) and predicted labels ($Y_{ni}$) for N observations and K classes. Since the classes are unbalanced, $w_i$ allows weighting.~\cite{mathworks_crossentropy_2024}
\begin{subequations}
    \label{eqn:crossentropy}
    \begin{equation}
        loss = \frac{1}{N} \sum^N_{n=1}\sum^K_{i=1} w_i*loss_i
    \end{equation}
    \begin{equation}
        loss_i = T_{ni}~ln(Y_{ni}) + (1 - T_{ni})~ln(1-Y_{ni})
    \end{equation}   
\end{subequations}

Two other loss functions that measure the overlap between the target and predictions - Jaccard index (Equation~\ref{subeq:jaccard}) and Dice similarity coefficient (Equation~\ref{subeq:diceloss}) were considered  while training the model, but the best-performing models were trained with cross-entropy loss. Y and T denote an image and its corresponding ground truth, K is the number of classes, M is the number of elements (pixels) in Y, $w_k$ is the class weight.
\begin{subequations}
    \begin{equation}
        J = \frac{Y~\bigcap~T}{Y~\bigcup~T}
        \label{subeq:jaccard}
    \end{equation}
    \begin{equation}
        S = \frac{2\sum^K_{k=1}w_k\sum^M_{m=1}Y_{km}T{km}}{\sum^K_{k=1}w_k\sum^M_{m=1}Y^2_{km}~+~T^2_{km}}
        \label{subeq:diceloss}
    \end{equation}
\end{subequations}

The best model was determined by evaluating various metrics, including overall accuracy and per-class accuracy. While overall accuracy may be higher for certain models, the actual performance could be worse due to differences in distribution of sample's classes. Accuracy is not always the best measure for object segmentation problems, so errors such as Intersection-over-Union (IoU) which measures the overlap of predictions to labels were also monitored. We also used Precision, Recall, and F1 Score, which aided in understanding the model's performance. Precision measures how precise each prediction is while Recall measures whether each class was classified correctly. F1 score is a measure of both Precision and Recall.
In Table~\ref{tab:bestresults}, 2D SegNet has the Mean IoU value in the Precision column and Mean BF score in the F1 Score column.

\begin{table}[htb]
    \centering
    \begin{ruledtabular}
    \begin{tabular}{c||cccc|ccc}
        & \multicolumn{4}{c|}{\underline{~Accuracy~}} & \multicolumn{3}{c}{\underline{~Other~}} \\
        % \hline
        Model & Overall & Substrate & Monolayer & Multilayer & Precision & Recall & F1 Score \\
        \hline\hline
        $\mathbf{2D SegNet}$ & $\mathbf{0.915}$ & $\mathbf{0.915}$ & $\mathbf{0.908}$ & $\mathbf{0.917}$ & $\mathbf{0.631}$ & NA & $\mathbf{0.503}$ \\ 
        $\mathbf{2D U-Net}$ & 0.960 & 0.970 & 0.737 & 0.758 & 0.721 & 0.654 & 0.665 \\
        $\mathbf{1D U-Net}$ & 0.933 & 0.942 & 0.000 & 0.524 & 0.526 & 0.503 & 0.500 \\
    \end{tabular}
    \end{ruledtabular}
    \caption{Best training results for each model.}
    \label{tab:bestresults}
\end{table}

\subsection{SegNet MATLAB}
SegNet is a deep convolution neural network with an encoder-decoder architecture, utilizing max-pooling to up-sample in the decoder stage, reducing the model size as previous models utilized fully connected networks to up-sample~\cite{han_deeplearningenabled_2020}.

For encoding and decoding, the process involves three layers of convolution, batch normalization, and ReLU activation, followed by either a max-pooling or max-unpooling layer, respectively. The depth of the network depends on how many times it is down-sampled (equivalent to the number of times it is up-sampled), culminating in a softmax classification layer for pixel-wise classification. The primary objective is to minimize pixel variation within the same class while maximizing differentiation between different classes, thereby ensuring precise boundary delineation.
Two critical parameters influencing this accuracy are the network depth (referring to the number of convolutional layers in both the encoder and decoder networks) and class weights \cite{badrinarayanan_segnet_2017}. SegNet's structure favors the predominant class (e.g., background substrate) if there's a disproportionate number of pixels in each class. To rectify this, smaller classes must be appropriately weighted in the loss function using a class weights array at the final pixel classification layer \cite{nanomaterial_deeplearning}.

As previously discussed, the Jaccard index, also known as Intersection over Union (IoU), measures the percentage of overlap between expected and predicted class areas. When evaluating visual results, the average IoU across all three classes proved to be the most effective metric for comparing different implementations using the same original dataset. For assessing implementations with altered parameters, monolayer accuracy emerged as the primary metric.
Less effective metrics included Weighted IoU, which averages IoU weighted by the number of pixels in each class, and Mean Boundary F1 Score, which assesses how well predicted boundaries match the correct boundaries. The variability of Mean Boundary F1 Score is illustrated in Tables~\ref{tab:2dsegnetparta} \& \ref{tab:2dsegnetpartb}, while the weighted average undermines the intent to appropriately handle larger class sizes.
Visual comparisons were conducted by delineating predicted or labeled monolayer boundaries in green and multilayer boundaries in blue on the original images.

Two main categories of implementations were tested. The baseline model was the already trained model taken from the GitHub repository~\cite{nanomaterial_deeplearning}.
\,Initially, the baseline model was compared to implementations that used transfer learning and the newly labeled dataset. First the model was trained with the original three classes: monolayer, multilayer, and background substrate. And the rest of the models used the fourth class of shear as it visually performed better as can be seen in Table~\ref{tab:2dsegnetparta}. The metrics could not capture the added class confusion in the shear region which would make the future implementations more confusing to visually compare. It does seem like the model still worked well in the relevant area. There were a large number of false positives in the normalized implementation so randomized cropping further improved the results as shown in Table~\ref{tab:2dsegnetparta}. Additionally, the various color preprocessing methods were compared but did not perform well according to the metrics nor visually. 

\begin{table}[htb]
    \centering
    \begin{ruledtabular}
     
    \begin{tabular}{||l||cccc|cc||}
        & \multicolumn{4}{c|}{\underline{~Accuracy~}} & \multicolumn{2}{c||}{\underline{~Other~}} \\
        Preprocessing & Overall & Substrate & Monolayer & Multilayer & Mean IoU & Mean BF  \\
        \hline\hline
        $\mathit{BASELINE}$ & $\mathit{0.855}$ & $\mathit{0.859}$ & $\mathit{0.764}$ & 
        $\mathit{0.875}$ & $\mathit{0.595}$ & $\mathit{0.346}$ \\
        \hline\hline\hline
        Normalized, 3~classes & 0.912 & 0.943 & 0.836 & 0.705 & 0.616 & 0.463 \\
        Normalized & 0.889 & 0.904 & 0.671 & 0.561 & 0.602 & 0.487 \\
        $\mathbf{Normalized,}$ $\mathbf{Random Crop}$ & $\mathbf{0.885}$ & $\mathbf{0.867}$ & $\mathbf{0.787}$ & $\mathbf{0.890}$ & $\mathbf{0.682}$ & $\mathbf{0.441}$ \\
        TCD:HSV & 0.825 & 0.813 & 0.528 & 0.698 & 0.535 & 0.407 \\
        TCD:Plasma & 0.821 & 0.790 & 0.459 & 0.800 & 0.539 & 0.393 \\
        TCD:Spectral & 0.857 & 0.834 & 0.338 & 0.827 & 0.570 & 0.430 \\
    \end{tabular}
    \end{ruledtabular}
    \caption{The metrics for the various implementations of the 2D Segnet model using transfer learning. The normalized preprocessing with random crop of the original images worked the best.}
    \label{tab:2dsegnetparta}
\end{table}

\begin{table}[htb]
    \centering
    \begin{ruledtabular}
    \begin{tabular}{||c|c||cccc|cc||}
        & & \multicolumn{4}{c|}{\underline{~Accuracy~}} & \multicolumn{2}{c||}{\underline{~Other~}} \\
        Transfer Learning & Preprocessing & Overall & Substrate & Monolayer & Multilayer & Mean IoU & Mean BF \\
        \hline\hline
        \multirow{4}{2em}{No} & None & 0.837 & 0.849 & 0.732 & 0.729 & 0.504 & 0.424 \\
        & TCD:HSV & 0.772 & 0.791 & 0.158 & 0.788 & 0.380 & 0.463 \\
        & TCD:Plasma & 0.799 & 0.822 & 0.132 & 0.792 & 0.389 & 0.294 \\
        & TCD:Spectral & 0.755 & 0.771 & 0.233 & 0.764 & 0.374 & 0.266 \\
        \hline
        \multirow{4}{1em}{Yes} & $\mathbf{None}$ & $\mathbf{0.915}$ & $\mathbf{0.915}$ & $\mathbf{0.908}$ & $\mathbf{0.917}$ & $\mathbf{0.631}$ & $\mathbf{0.503}$ \\
        & TCD:HSV & 0.809 & 0.830 & 0.310 & 0.752 & 0.406 & 0.380 \\
        & TCD:Plasma & 0.774 & 0.77 & 0.613 & 0.826 & 0.423 & 0.345 \\
        & TCD:Spectral & 0.790 & 0.800 & 0.511 & 0.772 & 0.413 & 0.351 \\
    \end{tabular}
    \end{ruledtabular}
    \caption{The metrics for the implementations of the 2D SegNet model using the prelabeled dataset with 1400 images. It compares the use of transfer learning to the dataset alone and compares some color processing options on the labeled images. }
    \label{tab:2dsegnetpartb}
\end{table}

\begin{figure}[htb]
    \centering
    \includegraphics[width=0.9\textwidth]{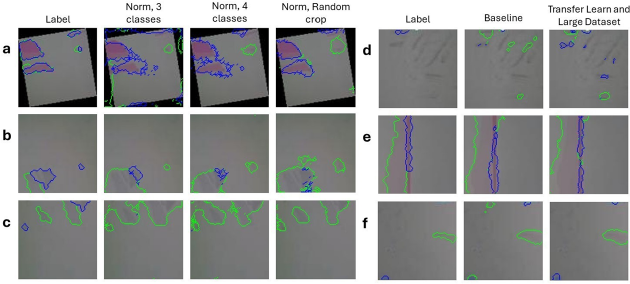}
    \vspace{-15pt}
    \caption{ A comparison of images predicted with three classes, four classes and random crop. (a)-(c) With our own 408 picture size dataset and (d)-(f)from the 1400 sized dataset that was prelabeled. Random crop performs better for (a-c) and transfer learning with large dataset performs best overall.}
    \label{fig:visual2dsegnet}
\end{figure}

The visual results for normalized, random crop can be seen in Figs.~\ref{fig:visual2dsegnet}a-c. However, the results are still not optimal which suggests that the dataset was too small. The models were then trained with a dataset of 1400 images that were prelabeled and accessible from the referenced GitHub. The performance of using transfer learning and not was compared. Table~\ref{tab:2dsegnetpartb} shows that the use of transfer learning with normalized color preprocessing performed best as can be confirmed in Fig~\ref{fig:visual2dsegnet}. 

\subsection{2D U-Net}
\begin{figure}[htb]
    \centering
    \includegraphics[width=0.5\textwidth]{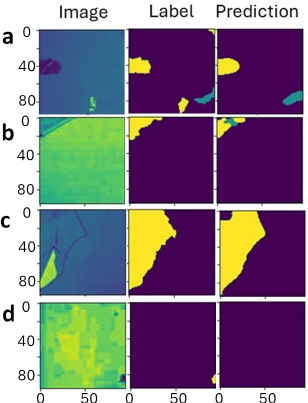}
    \vspace{-15pt}
    \caption{Labels and predictions using 2D U-Net for four microscopic images on different thickness in (a)-(d). Since python was used, the images cannot be transposed. Yellow is multilayer and green is monolayer. The axis is labeled for spatial position. }
    \label{fig:visual2dunet}
\end{figure}

U-Net is a convolution neural network with an encoder-decoder architecture, similar to SegNet, but also utilizes copy-and-crop operations on certain feature maps, creating a U-shape~\cite{ronneberger_unet_2015}. 2D U-Net models were trained on the full images and labels. Different preprocessing and hyperparameter tuning were used to improve the model's performance. Ultimately, the raw images normalized were found to produce the best performance. This may be due to the loss of certain features when moving to different color maps. One area of improvement would be to utilize the different channels between the color mappings rather than normalizing it to one value. This would increase the model's size, but potentially improve the model's accuracy. The results are listed in Table~\ref{tab:2dunet} and can be visualized in Fig.~\ref{fig:visual2dunet}.

\begin{table}[htb]
    \centering
    \begin{ruledtabular}
    \begin{tabular}{||c||cccc|ccc||}
        & \multicolumn{4}{c|}{\underline{~Accuracy~}} & \multicolumn{3}{c||}{\underline{~Other~}} \\
        Preprocessing & Overall & Substrate & Monolayer & Multilayer & Precision & Recall & F1 Score \\
        \hline\hline
        $\mathbf{None}$ & $\mathbf{0.960}$ & $\mathbf{0.970}$ & $\mathbf{0.737}$ & $\mathbf{0.758}$ & $\mathbf{0.721}$ & $\mathbf{0.654}$ & $\mathbf{0.665}$ \\
        Contrast & 0.871 & 0.908 & 0.016 & 0.064 & 0.330 & 0.334 & 0.320 \\
        TCD:Plasma & 0.953 & 0.963 & 0.473 & 0.716 & 0.633 & 0.590 & 0.593 \\
        TCD:HSV & 0.944 & 0.969 & 0.520 & 0.575 & 0.608 & 0.614 & 0.594 \\
        TCD:Spectral & 0.953 & 0.966 & 0.399 & 0.684 & 0.625 & 0.603 & 0.596 \\
    \end{tabular}
    \end{ruledtabular}
    \caption{Metrics for the different types of color pre-processing used for the 2D U-Net Model.}
    \label{tab:2dunet}
\end{table}

\subsection{1D U-Net}
\begin{figure}[htb]
    \centering
    \includegraphics[width=0.5\textwidth]{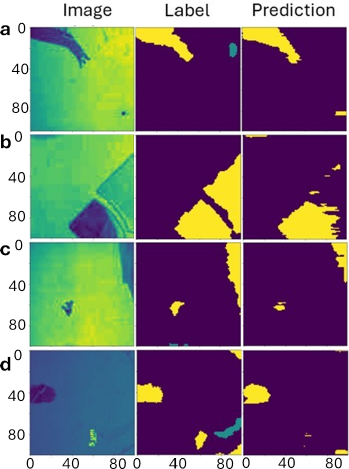}
    \vspace{-15pt}
    \caption{Labels and predictions using 1D U-Net for four microscopic imaging on different thickness (a)-(d). Using the same color scheme as 2D U-net where yellow is multilayer but green is monolayer. None of the predictions depict monolayers. Because they are generated line by line, the representation of the area is inadequate, resulting in regions that appear more streaky, as clearly seen in (b). The axis is labeled for spatial position.}
    \label{fig:visual1dunet}
\end{figure}

Each row in each image was utilized in training. During inferencing, the 100x100 image was separated into 100 rows, where each row was inferenced, and the resulting output row was appended to the resulting image's prediction. Hyperparameter tuning and different preprocessing was used to improve the model's performance. The results are displayed in Table~\ref{tab:1dunet} and can be visualized in Fig.~\ref{fig:visual1dunet}.

\begin{table}[htb]
    \centering
    \begin{ruledtabular}
    \begin{tabular}{||c||cccc|ccc||}
        & \multicolumn{4}{c|}{\underline{~Accuracy~}} & \multicolumn{3}{c||}{\underline{~Other~}} \\
        Preprocessing & Overall & Substrate & Monolayer & Multilayer & Precision & Recall & F1 Score \\
        \hline\hline
        None & 0.920 & 0.924 & nan & 0.501 & 0.500 & 0.453 & 0.450 \\
        $\mathbf{Contrast}$ & $\mathbf{0.932}$ & $\mathbf{0.943}$ & $\mathbf{NAN}$ & $\mathbf{0.524}$ & $\mathbf{0.526}$ & $\mathbf{0.503}$ & $\mathbf{0.500}$ \\
        TCD:Plasma & 0.923 & 0.930 & NAN & 0.454 & 0.502 & 0.461 & 0.458 \\
        TCD:HSV & 0.911 & 0.913 & NAN & 0.404 & 0.466 & 0.398 & 0.388 \\
        TCD:Spectral & 0.914 & 0.919 & NAN & 0.481 & 0.491 & 0.428 & 0.427 \\
    \end{tabular}
    \end{ruledtabular}
    \caption{Metrics for the different types of color pre-processing used for the 1D U-Net Model.}
    \label{tab:1dunet}
\end{table}

The 1D model did not perform as well as the 2D model, primarily due to the loss of information vertically. Another setback was the time to inference the model was longer than the 2D model, despite the model and input being smaller. One potential reason for this would be the movement of memory in the processor outweighing the efficiency of the model. Despite this, the model was able to get meaningful results.

\section{Conclusion and Outlook}
The best trained model classified the data classes in 2D, and encompassing all potential labels. Across all experiments, transfer learning consistently outperformed other methods, even when the pretrained model had no direct relevance to 2D materials, as seen with the U-Net model. Additionally, large datasets generally yielded superior performance compared to smaller, more precisely labeled datasets. However, differences in accuracy between datasets were more comparable when assessed with inferior models. Simply image normalization to the background proved more effective than any color processing method in most cases, except for one instance with the one-dimensional U-Net, possibly due to clearer differentiation between the background and monolayer/multilayer. 

This study was limited to a single dataset, restricting broader conclusions. However, the introduction of new labeling capabilities allows for future exploration into how dataset diversity correlates with model accuracy. Moreover, refining data augmentation methods, particularly shear augmentation, could potentially obviate the need for four distinct classes.

Future research should focus on leveraging image processing techniques to automate labeling by extracting contrast data for layer assignment. This involves normalizing images using total color difference to standardize images captured in different environments. Furthermore, expanding beyond monolayer and multilayer identification by iterating through the "define baseline labeling" step could reveal additional layers, typically up to a maximum of ten.
Lastly, model performance can be enhanced by fine-tuning hyperparameters and optimizing transfer learning through weight freezing strategies.

\newpage
\textbf{References}
\bibliographystyle{IEEEtran}
\bibliography{machine-learning}

\end{document}